\documentclass[12pt,preprint]{aastex}
\usepackage{psfig} 

\def\be{\begin{equation}}
\def\ee{\end{equation}}

\def\m{~$\mu$m}

\def\ISO{{\it ISO}}
\def\IRAS{{\it IRAS}}
\def\SIRTF{{\it SIRTF}}

\def\IRAScolor{${f_\nu (60 \mu {\rm m})} \over {f_\nu (100 \mu {\rm m})}$}

\def\colorb{${f_\nu (6.75 \mu {\rm m})} \over {f_\nu (850 \mu {\rm m})}$}
\def\colorc{${f_\nu (15 \mu {\rm m})} \over {f_\nu (850 \mu {\rm m})}$}
\def\colord{${f_\nu (850 \mu {\rm m})} \over {f_\nu (1.4 {\rm GHz})}$}

\begin {document}
\slugcomment{\scriptsize \today \hskip 0.2in Version 4.0}

\title{The Infrared Spectral Energy Distribution of Normal Star-Forming Galaxies: Calibration at Far-Infrared and Submillimeter Wavelengths}
 
\author{Daniel A. Dale}
\affil{Department of Physics and Astronomy, University of Wyoming, Laramie, WY 82071}
\and
\author{George Helou}
\affil{IPAC, California Institute of Technology 100-22, Pasadena, CA 91125}

\begin {abstract}
New far-infrared and submillimeter data are used to solidify and to extend to long wavelengths the empirical calibration of the infrared spectral energy distribution (SED) of normal star-forming galaxies.  As was found by Dale et al. (2001), a single parameter family, characterized by \IRAScolor, is adequate to describe the range of normal galaxy spectral energy distributions observed by \IRAS\ and \ISO\ from 3 to 100\m.  However, predictions based on the first generation models at longer wavelengths (122 to 850\m) are increasingly overluminous compared to the data for smaller \IRAScolor, or alternatively, for weaker global interstellar radiation fields.  After slightly modifying the far-infrared/submillimeter dust emissivity in those models as a function of the radiation field intensity to better match the long wavelength data, a suite of SEDs from 3\m\ to 20~cm in wavelength is presented.  Results from relevant applications are also discussed, including submillimeter-based photometric redshift indicators, the infrared energy budget and simple formulae for recovering the bolometric infrared luminosity, and dust mass estimates in galaxies.  Regarding the latter, since galaxy infrared SEDs are not well-described by single blackbody curves, the usual methods of estimating dust masses can be grossly inadequate.  The improved model presented herein is used to provide a more accurate relation between infrared luminosity and dust mass.
\end {abstract}
 
\keywords{dust, extinction --- galaxies: general --- galaxies: ISM --- infrared: galaxies --- infrared: ISM: continuum}

\section {Introduction}

A new model for the infrared spectral energy distribution of star-forming galaxies, based primarily on {\it Infrared Astronomical Satellite} (\IRAS) and {\it Infrared Space Observatory} (\ISO) observations of 69 normal galaxies (Dale et al. 2000), was developed by Dale et al. (2001).  A particularly useful feature to these models is that the various SED shapes are adequately described by a single-parameter family, namely depending only on \IRAScolor, the ratio of the two far-infrared flux densities measured by \IRAS\ and an indicator of the typical dust grain temperature; the full suite of galaxy infrared SEDS can be described as spanning a range of far-infrared colors, and thus dust temperature.  However, though they span a large wavelength range, 3-1100\m, the model curves were only constrained using data from 3-100\m.  This dearth of far-infrared and submillimeter data for normal star-forming galaxies is largely a consequence of prior instrumental limitations:  \IRAS\ did not probe beyond 100\m\ and the first-generation submillimeter instruments for continuum observations were single-channel broadband photometers, difficult for generating spatial maps of cold dust in large nearby galaxies.   

Fortunately, with the opportunities provided by \ISO\ and the advent of sensitive submillimeter detector arrays, there are now substantial far-infrared and submillimeter databases for normal star-forming galaxies.  For example, Dunne et al. (2000, 2001) have used the Submillimetre Common-User Bolometer Array, or SCUBA, to systematically map 114 galaxies from the \IRAS\ Bright Galaxy Sample (Soifer et al. 1989) at 450 and 850\m\ data.  Furthermore, a template for the far-infrared continuum of galaxies is being constructed, and is based on \ISO\ Long Wavelength Spectrometer continuum data for 228 galaxies at 52, 57, 63, 88, 122, 145, 158, and 170\m\ (Brauher 2002).  And Stickel et al. (2000) provide 170\m\ data from the ISOPHOT Serendipity Survey for 115 galaxies.  We use these data to constrain our SED model beyond 100\m.

Infrared continuum data are frequently used to estimate the dust mass of galaxies.  However, these calculations typically assume that the infrared emission arises from a single temperature blackbody.  Surely, the integrated infrared emission from galaxies stems from dust spanning a wide range of temperatures.  In fact, some of the dust emission is not even from dust in thermal equilibrium---polycyclic aromatic hydrocarbons (PAHs) and very small grains are stochastically heated (Draine \& Li 2001).  How do the `true' dust masses for the model SEDs compare with those derived from the canonical single-temperature blackbody approach?   As will be shown, the standard approach severely underestimates galaxy dust masses.

\section {A Phenomenological Model for the Infrared Spectral Energy Distributions}
\label{sec:model}

D\'{e}sert, Boulanger \& Puget (1990) developed infrared SEDs appropriate for dust emission from individual regions within the Milky Way.  Their models were based on three dust components: large ``classical'' grains that are in thermal equilibrium, semi-stochastically heated ``very small grains,'' and two-dimensional polycyclic aromatic hydrocarbons stochastically heated by individual ultraviolet or optical photons.  Starting from the framework laid out by D\'{e}sert, Boulanger \& Puget for Milky Way dust emission, Dale et al. (2001) constructed a wide range of semi-empirical infrared SEDs, with the shape and normalization of each determined by the intensity $U$ of the interstellar radiation field, normalized such that $U=1$ for the local interstellar radiation field; the models span $0.3\leq U \leq 10^5$.  They reasoned that since the {\it integrated} emission from normal galaxies results from a superposition of emission from these various environments, it is reasonable to compute ``global'' spectra from a power-law distribution of the local SEDs.  The local SEDs are combined assuming a power-law distribution in a given galaxy of dust mass over heating intensity:
\begin{eqnarray}
\label{eq:dMdU}
             dM_d(U) &\propto& U^{-\alpha} \; dU,
\end{eqnarray}
where $M_{\rm d}(U)$ is the dust mass heated by a radiation field at intensity $U$, and the exponent $\alpha$ is a parameter that represents the relative contributions of the different local SEDs.

The SED model of Dale et al. (2001) was constrained by \IRAS\ and ISOCAM broadband photometric and ISOPHOT spectrophotometric observations of a sample of 69 normal star-forming galaxies between 3 and 100\m\ ($L_{\rm FIR}$ from less than $10^8~L_\odot$ to as large as $10^{12}~L_\odot$).  The model reproduces well the empirical spectra and infrared color trends; the range of $\alpha$ that describes the suite of normal galaxy SEDs is approximately $1<\alpha<2.5$.  A power law distribution model for the radiation field was also successfully applied to the Small Magellanic Cloud SED (Li \& Draine 2002).  Interestingly, power-law profiles with similar exponents are found for the mass distributions of clouds in the interstellar medium (Stutzi 2001; Shirley et al. 2002; Elmegreen 2002).

\section{Comparison of Previous Model with Long Wavelength Data}

\subsection {Comparison with SCUBA 850\m\ Observations}

\begin{figure}[!ht]
\plotone{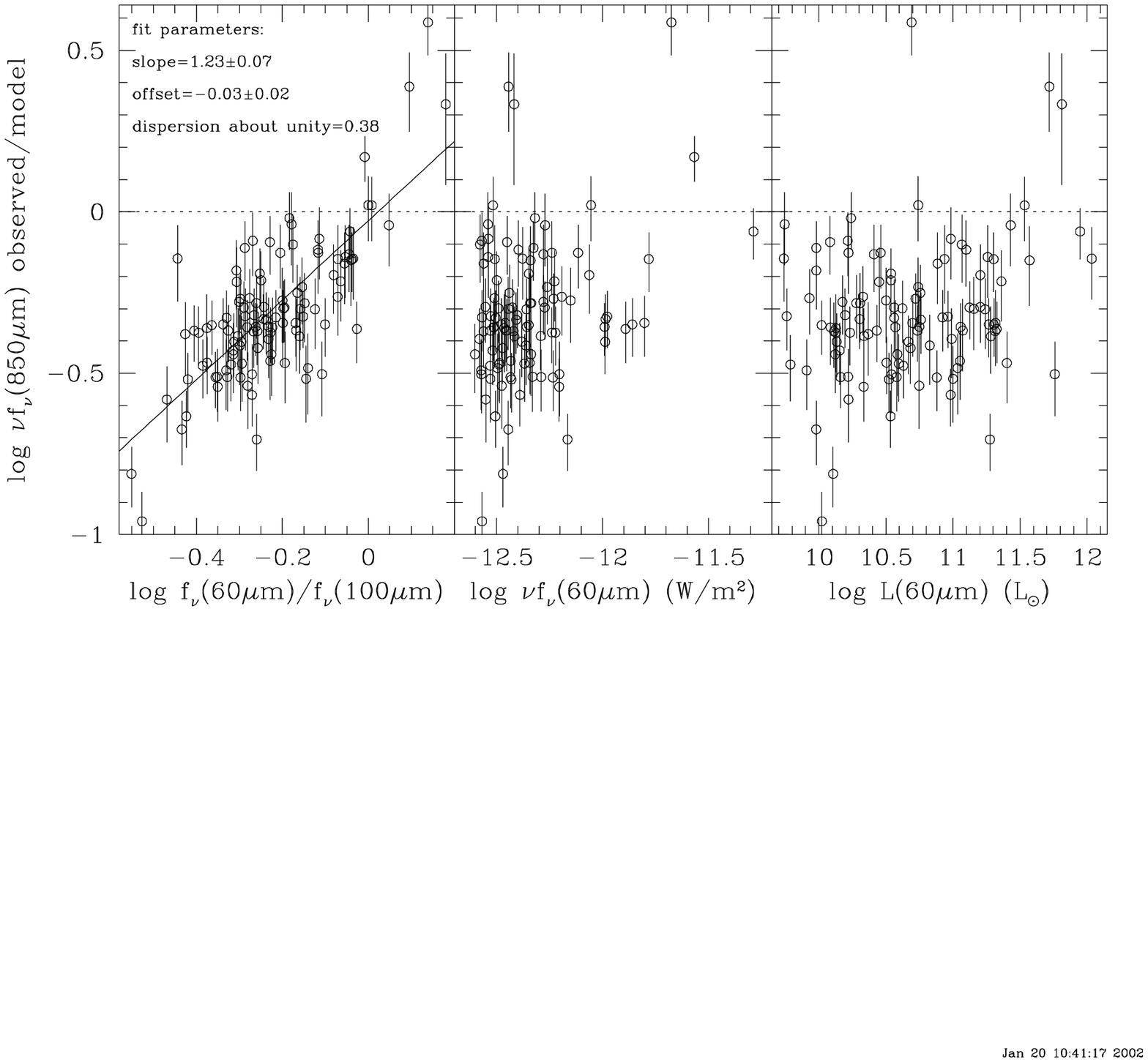}
\caption[] {\ A comparison of observed SCUBA 850\m\ fluxes (Dunne et al. 2000) with the 850\m\ fluxes predicted from the SED model of Dale et al. (2001).  Here the luminosity at 60\m\ is defined as $4\pi d^2 \nu f_\nu(60~\micron)$ where $d$ is the distance to the galaxy in the Local Group rest frame.} 
\label{fig:850_before}
\end{figure}

Figure~\ref{fig:850_before} shows a comparison of observed SCUBA 850\m\ fluxes for 110 galaxies (Dunne et al. 2000) with the 850\m\ fluxes predicted from the suite of SED models Dale et al. (2001) derived for normal star-forming galaxies.  For each SCUBA source, a corresponding SED model is selected solely according to \IRAScolor.  Then an 850\m\ model flux is generated by convolving the model SED with the SCUBA 850\m\ filter and atmospheric transmissivity profiles.  No trend is seen as a function of far-infrared flux and luminosity, but a clear trend is seen with far-infrared color: the model prediction is increasingly overluminous at 850\m\ for progressively cooler galaxies.  Moreover, the average model is overluminous by a factor of two.  One potential explanation would be an overestimation of dust emissivity at long wavelengths, which is an unconstrained parameter.

Are similar trends seen within galaxies?  This question is unfortunately difficult to answer since there are very few complete submillimeter maps of large nearby galaxies, in which the large angular extent of the dust emission allows several spatially-independent pointings.  Bianchi et al. (2000) have obtained such a map for NGC~6946.  Using their submillimeter data, a model bias similar to that seen in Figure~\ref{fig:850_before} is found within the disk of NGC~6946.

\subsection {Comparison with \ISO\ Far-Infrared Spectroscopy and Photometry}

The archival ISOLWS work of Brauher (2002) provides a useful resource for checking the far-infrared continuum levels of the SED model.  Part of his study concentrates on measuring galaxy far-infrared continuum levels at 52, 57, 63, 88, 122, 145, 158, and 170\m\ based on \ISO\ Long Wavelength Spectrometer data for 228 galaxies (183 of which are unresolved by the LWS beam).  After fitting and subtracting any line emission, Brauher gauges the baseline continuum level detected by \ISO\ over a 2 to 5\m-wide spectral window centered on each line wavelength.

Each of the galaxies studied by Brauher (2002) was detected by \IRAS\ at 60 and 100\m.  Since the models are parametrized by \IRAScolor\ the particular SED shape selected for each galaxy is based solely on the observed far-infrared color.  The predicted continuum level at each line wavelength is then scaled by the observed \IRAS\ flux density (the result is independent of whether $f_\nu(60~\micron)$ or $f_\nu(100~\micron)$ is used).  Figure~\ref{fig:fir_after} displays a comparison between the far-infrared continuum levels observed by ISOLWS (open circles) with those predicted by the models.  The outliers are the fainter galaxies with low (or possibly even non-detected) continuum levels that are particularly affected by any errors in dark current subtraction (as large as 10~Jy for some observations) and perhaps light leaks (J. Brauher, private communication).  Also portrayed in Figure~\ref{fig:fir_after} is a comparison involving the 170\m\ continuum levels from the ISOPHT Serendipity Survey (Stickel et al. 2000).  The trend is similar to what is seen using the 170\m\ data of Brauher (2002), and is in the same sense as the trend at 850\m: the models are overluminous for the coldest galaxies yet become progressively underluminous for galaxies with increasingly blue far-infrared colors.  

If the colder galaxies in the Stickel et al. and the Brauher samples subtend sizes larger than the instrument fields of view, and conversely if the hotter galaxies are compact enough to fit within the beams, then the observed trends could result.  As stated above, 45 of the 228 galaxies from the Brauher study are resolved by LWS and thus the observed continuum fluxes may be underestimated due to missing flux from the galaxy outskirts.  However, this is unlikely to be a major concern for these galaxies since the 60 and 100\m\ fluxes extracted from the ISOLWS spectra agree with the global \IRAS\ fluxes to within 5\%.  The trend in the data from the Serendipity Survey is likely not a remnant of beam size effects or their data reduction techniques since the envelope of their {\it photometric} data agrees with the envelope of Brauher's {\it spectroscopic} data.  Moreover, the ISOPHOT data from Stickel et al. derive from a much larger beam than the Brauher ISOLWS data.  The best agreement between the far-infrared observations and the models holds for $\lambda<100$\m, but this is not surprising given that the SED model of Dale et al. (2001) was based on existing data from 3 to 100\m.  The combination of good agreement at $\lambda<100$\m\ and increasing overestimation at longer wavelengths again points to a possible overestimation of dust emissivity at longer wavelengths.

\section {An Improved Description of Dust Emissivity}

\subsection {Evidence for Variations in Dust Emissivity}

An important factor in modeling the far-infrared emission from dust is the emissivity index $\beta$, a parameter that relates a dust grain's emissivity to the wavelength of emission: $\epsilon_\nu \propto \nu^{-\beta}$.  Typical reported values for dust emissivity range from $\beta=1$ to $\beta=2$, but anecdotal evidence supports an emissivity parameter that depends on environment and that may reach more extreme values.  One line of evidence for an environmentally-sensitive $\beta$ comes from the Dunne et al. (2000) SCUBA data: $\beta$ appears to decrease with increasing dust temperature $T_{\rm dust}$.  Additional 850\m\ SCUBA data imply that $\beta\sim1$ for luminous infrared galaxies with higher infrared-to-blue ratios (see Figure 2 of Lisenfeld, Isaak \& Hills 2000).  On the other hand, balloon-based submillimeter observations of interstellar clouds indicate that $\beta$ can reach quite large values.  Data from the Polaris flare and of a translucent filament in the Taurus molecular complex show that $\beta=1.9-2.2$ for extremely cold ($12-13$~K) dust (Bernard et al. 1999; Stepnik et al. 2001).  Recent findings by Dupac et al. (2001) support these claims.  They find from similar balloon-based submillimeter observations of Orion and M17 molecular cloud regions that $\beta$ varies from 1 to 2.5 and is anticorrelated with the local dust temperature: the larger values correspond to the colder locales ($\sim15$~K).  Furthermore, star counts towards Polaris and Taurus suggest that the cold dust component far-infrared/submillimeter emissivity is about three times higher than for the warm dust (Cambr\'{e}sy et al. 2001; Stepnik et al.).  
Finally, work by Pollack et al. (1994) shows that $\beta$ can be as large as 2.5 in the far-infrared/submillimeter for molecular clouds and accretion disks, but may drop to near zero at longer (millimeter) wavelengths.  Clearly, there is a growing body of observational evidence that $\beta$ can vary substantially, and that it is correlated with the characteristics of the overall local environment.

Laboratory studies also indirectly point to a similar scenario.  Work on different amorphous silicates in the submillimeter/millimeter wavelength range shows that $\beta$ ranges from 1.2 to 2.7 (Agladze et al. 1996), and is anticorrelated with dust temperature.  This trend holds down to $T_{\rm dust}\sim$ 10~K, but Agladze et al. did not probe to wavelengths shorter than the submillimeter.  Menella et al. (1998), however, did study crystalline and amorphous grains from 20\m\ to 2~mm, and they find a similar anticorrelation, over a temperature range $T_{\rm dust}=24$ to 295~K.

\subsection{Variable Dust Emissivity Model}
\label{sec:beta}

\begin{figure}[!ht]
\plotone{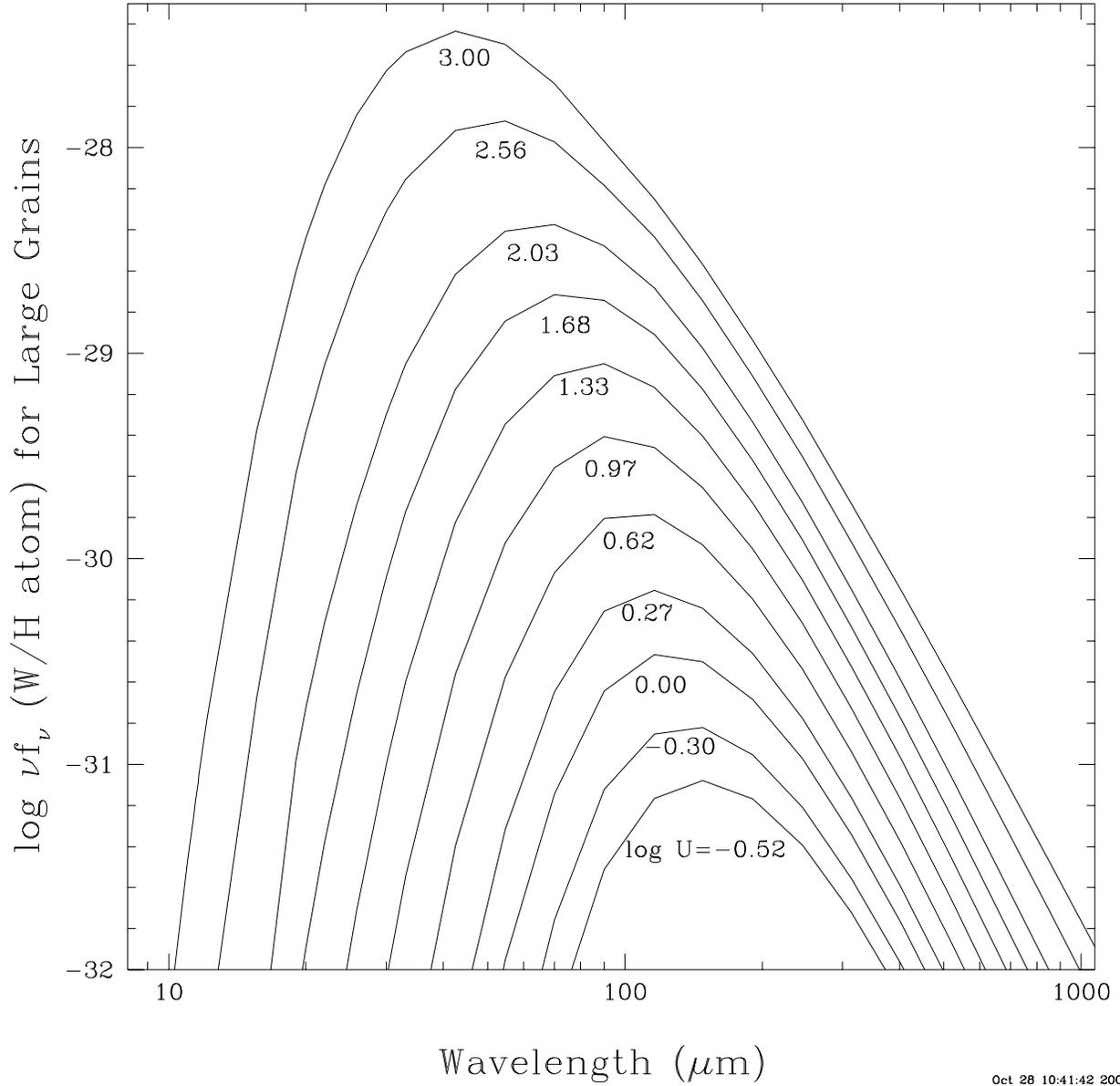}
\caption[] {\ Sampling of the collection of large grain emission profiles for heating intensities ranging from 0.3 to $10^3$ times the local interstellar radiation field.  The curves are computed incorporating the variable dust emissivity index outlined in Section~\ref{sec:beta} and described by Equation~\ref{eq:beta_tweak}.}  
\label{fig:large_grains}
\end{figure}

\begin{figure}[!ht]
\plotone{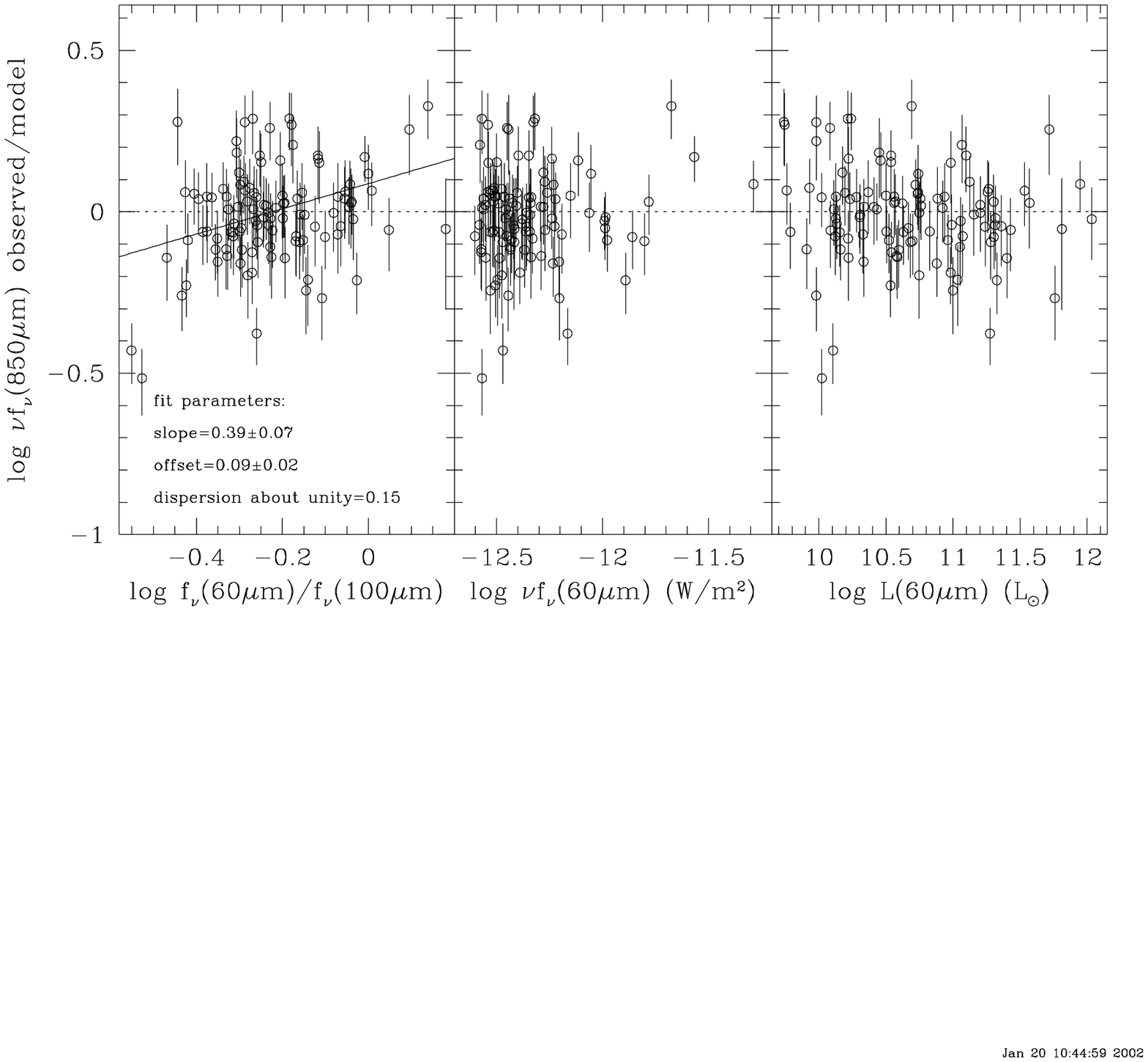}
\caption[] {\ Similar to Figure~\ref{fig:850_before}, but after invoking the variable far-infrared dust emissivity outlined in Section~\ref{sec:beta} and quantified by Equation~\ref{eq:beta_tweak}.} \label{fig:850_after}
\end{figure}

\begin{figure}[!ht]
\plotone{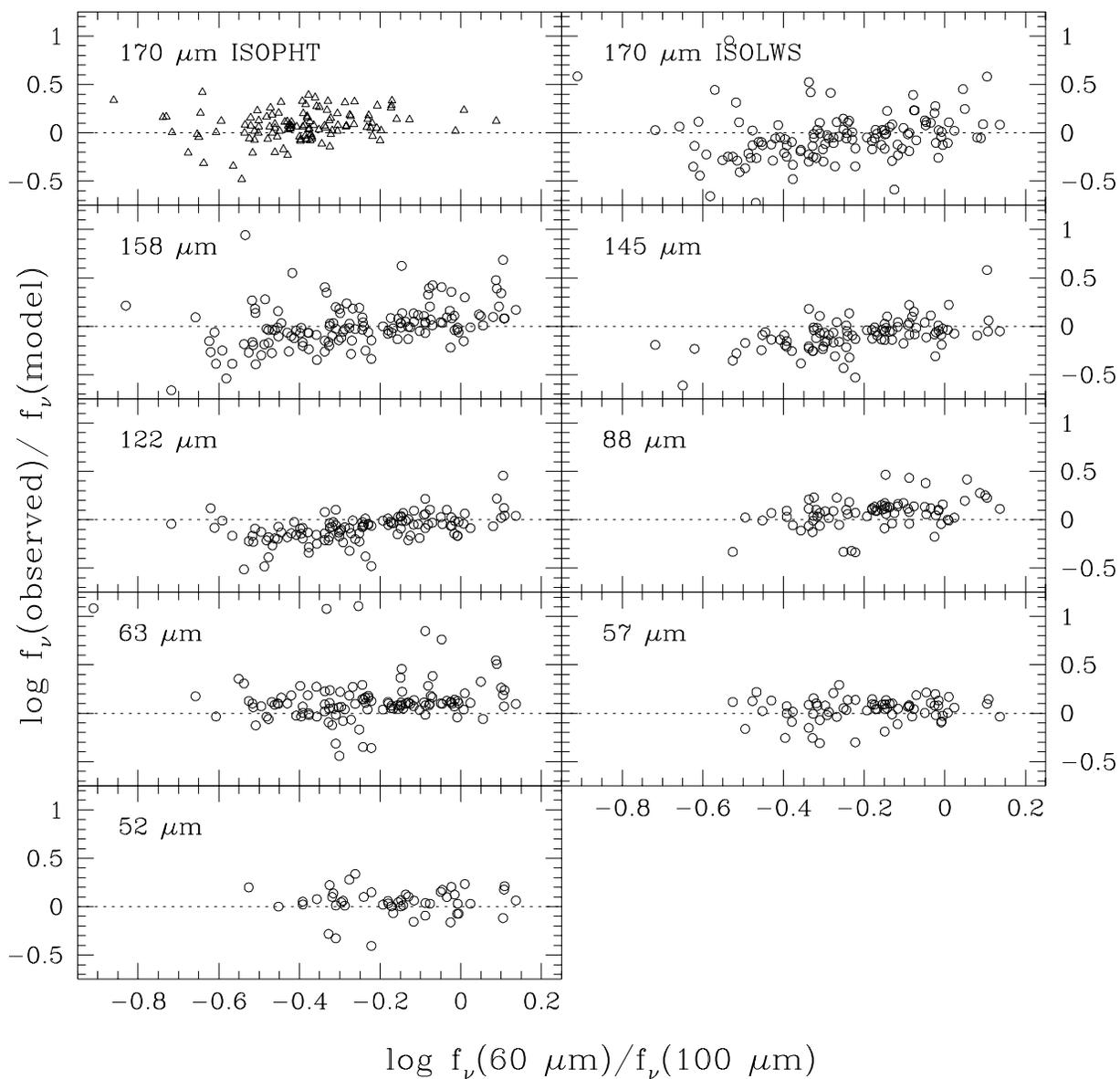}
\caption[] {\ Comparison of observed far-infrared continuum levels observed by \ISO\ with the model predictions.
The circles derive from the ISOLWS template work of Brauher (2002) and the triangles represent 170\m\ data from the ISOPHT Serendipity Survey (Stickel et al. 2000).}  
\label{fig:fir_after}
\end{figure}

In the original work of D\'{e}sert, Boulanger \& Puget (1990) and continued in Dale et al. (2001), large grains are responsible for the bulk of the far-infrared emission of galaxies.  The far-infrared emissivity of this material was assumed to follow an index $\beta=1.5$ for $20~\micron<\lambda<100~\micron$ and $\beta=2.0$ beyond 100\m.  In keeping with the philosophy that infrared SED shapes are to first order a function of $U$ (or equivalently \IRAScolor), we correct for the residuals in Figure~\ref{fig:850_before} according to $U$.  One such remedy is to alter the large grain dust emissivity parameter:
\be
\beta = 2.5 - 0.4 \log U.
\label{eq:beta_tweak}
\ee
Since previously the SED model was empirically calibrated with data at $\lambda < 100$\m, this new modification is only invoked for $\lambda>100$\m.  This relation generally follows the trend suggested by the data: $\beta$ increases (dust emissivity decreases) as the heating intensity decreases.  For $[U=1,10,1000]$, the index is $[\beta=2.5,2.1,1.3]$.  Figure~\ref{fig:large_grains} displays the large grain emission profiles for the range $U=0.3-10^3$ and Figure~\ref{fig:850_after} gives the resulting model-data residuals at 850\m.  Equation~\ref{eq:beta_tweak} was selected based on a simple optimization criterion---it minimizes the data-model residuals.  Note that in addition to having removed the obvious trend with \IRAScolor, there continue to be no significant trends as a function of flux and luminosity and the overall ratio means are now centered on unity.  The slope of the linear fit to the residuals in the lefthand panel has dropped from 1.23 to 0.39 (see also Figure~\ref{fig:850_before} and Table~\ref{tab:fits}).  

The far-infrared residuals portrayed in Figure~\ref{fig:fir_after} also decrease after invoking the variable dust emissivity index prescribed in Equation~\ref{eq:beta_tweak}.  The trends in the far-infrared residuals can be essentially completely removed with a more drastically variable dust emissivity index, $\beta = 2.8 - 0.8 \log U$, but then this overcompensates for the initial trend seen at 850\m.  Given that the 850\m\ data are more reliable, we shall proceed with the prescription outlined by Equation~\ref{eq:beta_tweak}.  The relative improvement in the far-infrared residuals is described in Table~\ref{tab:fits}.  The improvement is larger at longer wavelengths because the change in dust emissivity is only applicable for $\lambda>100$\m; the trends change a bit for $\lambda<100$\m\ since the predicted \IRAS\ 100\m\ fluxes are slightly different, and thus the \IRAScolor-based model predictions are slightly influenced.

\section{Results}

\subsection{Presentation of new SEDs}
\label{sec:new_seds}

\begin{figure}[!ht]
\plotone{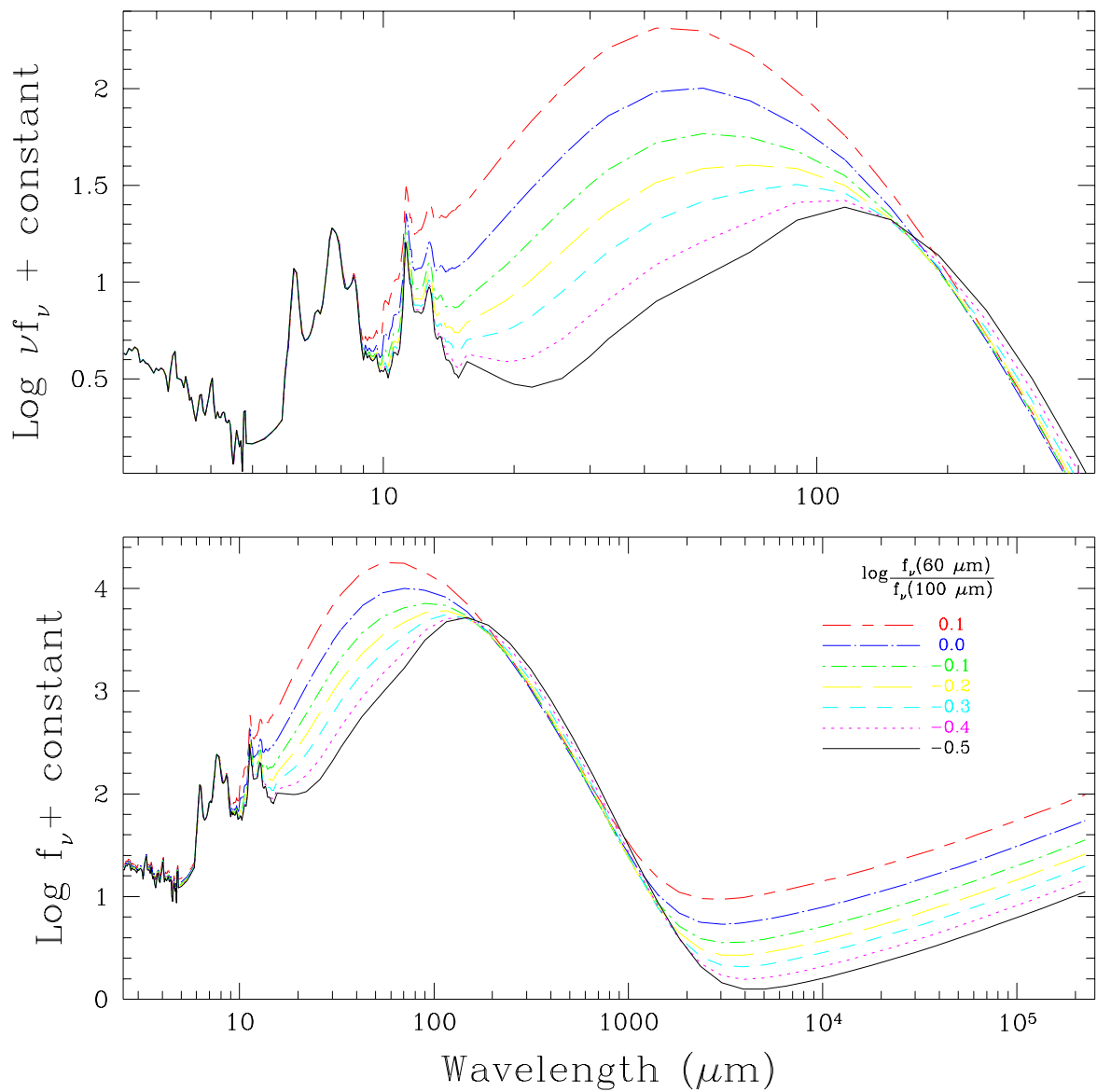}
\caption[] {\ Sampling of the galaxy model spectra, arbitrarily normalized at the 6.2\m\ feature.  The infrared portion of the model SEDs are displayed in the upper panel.  The extension to radio wavelengths is portrayed in the lower panel.  Note that the abscissa in the lower panel is $f_\nu$ whereas the upper panel gives $\nu f_\nu$.}  
\label{fig:seds}
\end{figure}

Figure~\ref{fig:seds} shows the updated infrared SEDs.  As portrayed in the lower panel of Figure~\ref{fig:seds}, the curves have also been extended to radio wavelengths.  The extension relies on the well-known far-infrared-to-radio correlation for galaxies: $q=\{\log \left[ {{\rm FIR} \over 3.75 \times 10^{12}~{\rm Hz}}\right] / f_\nu(1.4~{\rm GHz})$\}, where ``FIR'' is the \IRAS-based definition for the 42-122\m\ far-infrared flux (see Helou, Soifer \& Rowan-Robinson 1985; Helou et al. 1988).  A recent analysis of \IRAS\ and radio continuum data for 1809 galaxies by Yun, Reddy \& Condon (2001) gives $q \simeq 2.34\pm0.01$ with a 1$\sigma$ dispersion of 0.26~dex.  The model SEDs assume a constant ratio $q$, as there is no correlation with either far-infrared color or luminosity (Yun, Reddy \& Condon; and M. Yun, private communication).  The radio flux arises from two general components:  non-thermal processes such as synchrotron radiation resulting from the interaction of supernova-generated cosmic rays and interstellar magnetic fields, and thermal free-free emission from ionized gas (Helou \& Bicay 1993).  As outlined by Condon (1992), the thermal fraction of the flux density at 20~cm in normal star-forming galaxies is about 10\% and exhibits a rather flat spectral slope of $f_\nu^{\rm th} \propto \nu^{-0.1}$; 90\% is non-thermal and typically $f_\nu^{\rm non-th} \propto \nu^{-0.8}$.  These fractions and spectral indices are employed in the model.

\subsection{Improving Upon Standard Dust Mass Estimations}

The large dust grains in normal galaxies are assumed to be in thermal equilibrium, and thus the dust mass can be computed for a given flux density $f_\nu$ if the blackbody temperature $T_{\rm d}$ is known or estimated:
\be
M_{\rm dust} \propto {f_\nu \over \kappa_\nu B_\nu(T_{\rm d})}
\label{eq:Md}
\ee
where $\kappa_\nu \propto \nu^\beta$ is the mass absorption coefficient (e.g. Li \& Draine 2001).  Observers typically estimate galaxy dust masses using this technique, and determine the dust temperature by fitting one (or perhaps two or three) blackbodies to a small number of far-infrared and submillimeter fluxes.  However, this approach assumes that the large grains in all environments within a galaxy are heated to the same dust temperature, and that they all exhibit the same dust emissivity properties.

Our ``global'' SED model allows a more realistic derivation of the dust mass, because it combines SEDs from the full range of heating environments.  Moreover, since the large grains in the local SED models are in thermal equilibrium, Equation~\ref{eq:Md} can appropriately be used to compute the local large grain contribution.  Then Equation~\ref{eq:dMdU} may be used to determine the actual global large grain mass.  The PAHs and very small grains together contribute, at most, an additional 14\% to the total dust mass in our scheme (see Section~5.4 in Dale et al. 2001).

\begin{figure}[!ht]
\plotone{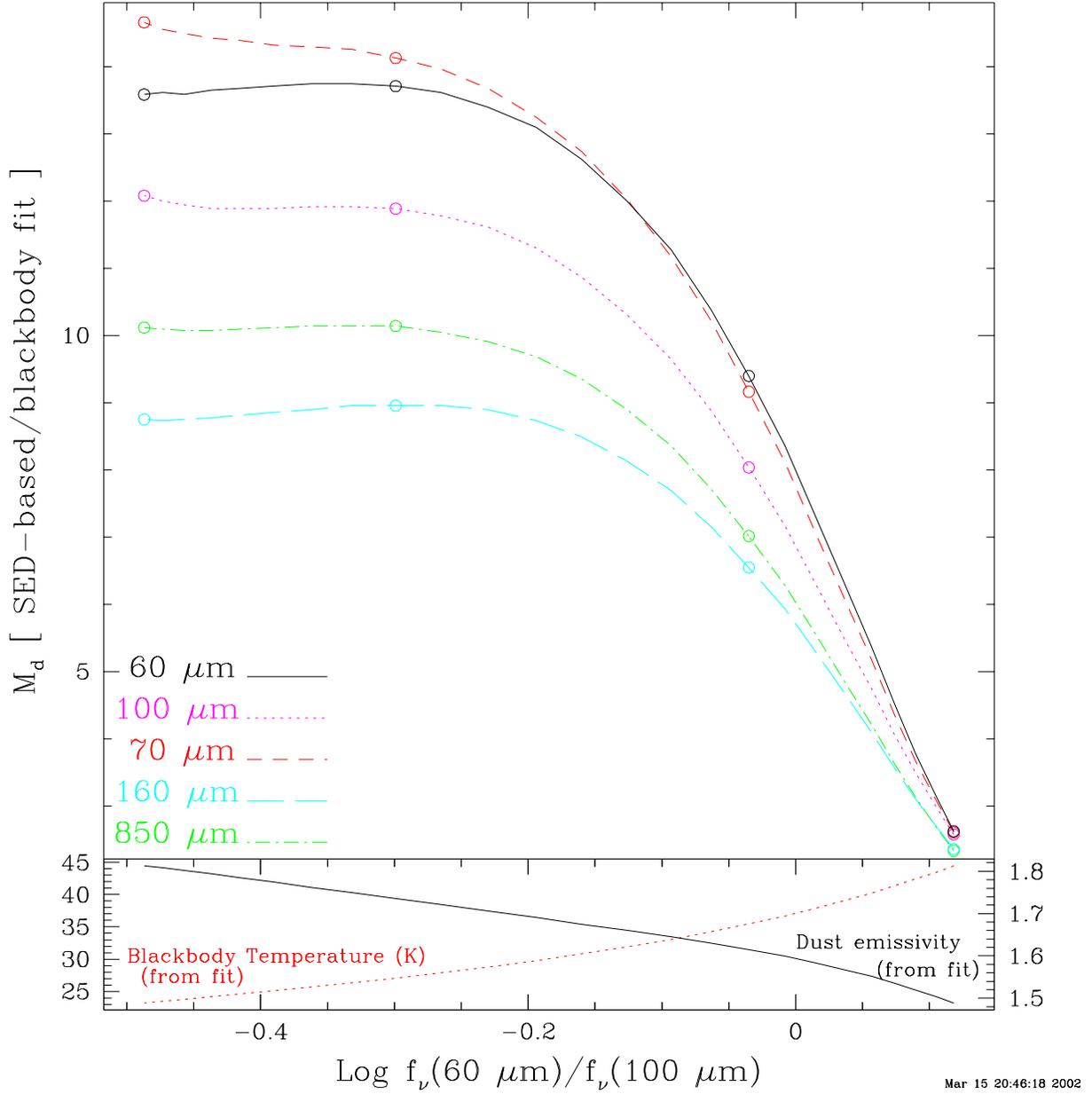}
\caption[] {\ The upper panel shows how dust masses computed from single blackbody fits to the infrared SEDs compare to the actual dust masses.  The various curves correspond to using broadband flux densities at different wavelengths.  The dotted (solid) curve in the lower panel gives the best-fit blackbody temperature (dust emissivity index) for each SED.  The sequence of open circles indicates models that correspond to power law exponent values of $\alpha=$[2.5,2.0,1.5,1.0] (in order of increasing \IRAScolor).}
\label{fig:dust_mass}
\end{figure}

The upper panel of Figure~\ref{fig:dust_mass} displays the ratio of the dust mass from the SED model to that predicted from a single modified blackbody approximation for the SED (again using Equation~\ref{eq:Md}).  The blackbody temperature and dust emissivity are estimated from a fit to the simulated \IRAS\ 100\m, \SIRTF\ 70 and 160\m\ and SCUBA 850\m\ fluxes and displayed in the lower panel.  The dust mass is then estimated using the single blackbody fit and the flux at each wavelength as labelled in Figure~\ref{fig:dust_mass}.  It is clear from the figure that using a single blackbody provides a significant underestimate of the dust mass, even for the most active galaxies for which the undersampling factor is ``only'' a few.  The reason for this large discrepancy lies in the single blackbody approximation: the global SEDs are constructed from a combination of blackbody emission profiles that span a wide range of temperatures and emissivities; the various blackbody profiles significantly differ, especially in the far-infrared.  Moreover, the very small grain emission, and to a lesser extent the emission from PAHs, affect the shape of the far-infrared SED at 70\m\ (see Figure~5 of Dale et al. 2001) and thus have an impact on the far-infrared flux densities.  

The reason the ratios displayed in Figure~\ref{fig:dust_mass} trend toward smaller values for the more active galaxies is easily explicable: for such galaxies the effective spread in the radiation field intensity $U$ shrinks--the strong contributions from the hot, large grains in thermal equilibrium dominate the overall SED shape.  However, it may seem a bit strange that the ratios in Figure~\ref{fig:dust_mass} are a function of wavelength---Why should the dust mass depend on wavelength?  As alluded to above, the SED-based dust masses stem from a power law combination of blackbodies over a large range of dust temperature, whereas the blackbody-based dust masses are derived from a single fit to the broadband fluxes.  The discrepancy between the single blackbody fit and the model SED changes with wavelength, thus giving rise to the range of ratios in Figure~\ref{fig:dust_mass}.

\subsection{The Infrared Energy Budget}

Table~\ref{tab:budget} provides the infrared energy budget, updated from the version in Dale et al. (2001) to include the modifications to the far-infrared/submillimeter dust emissivity described in Section~\ref{sec:beta}.  The table shows how much energy emerges in various infrared bands for model galaxies with different star formation activity, parametrized by the \IRAS\ \IRAScolor\ far-infrared color in the first column.  The headings for columns 3--8 give the wavelength range over which the spectrum is integrated, and the table entries are the fraction of total infrared luminosity appearing in that range.  The numbers are largely the same as those presented in Dale et al. (2001), except for the submillimeter range.  After modifying the far-infrared/submillimeter dust emissivity, the 122--1100\m\ flux fraction now spans 6--32\%, whereas the previous percentage range was 5--39\%.  The models suggest that the far-far-infared and submillimeter flux from even the coldest normal galaxies is no greater than one-third of the total infrared flux.

The spectral range in column 7 corresponds to the FIR synthetic band (\S~\ref{sec:new_seds}), a commonly quoted infrared flux quite often used as an indicator of the total level of star formation activity.  This particular wavelength range accounts for about half of the total infrared emission for the majority of normal galaxies (i.e. those with \IRAScolor\ $\gtrsim0.5$).  Another interesting feature to the infrared energy budget is the relative constancy of the mid-to-total-infrared ratio, suggesting a tight link between the global star formation rate and the emission from the smallest interstellar grains.  The utility of the mid-infrared flux is explored further in \S~\ref{sec:mir_submm}. 

Simple relations that derive the far-infrared or bolometric infrared flux using a limited number of standard filters have wide applications.  It is generally believed (e.g. Mann et al. 2002) that the integrated 3--1000\m\ infrared luminosity best reproduces model star formation rates.  The FIR 42-122\m\ flux mentioned above is based on the \IRAS\ 60 and 100\m\ flux densities and is applicable to most galaxies.  Sanders \& Mirabel (1996) have also derived a relation for the $8-1000$\m\ bolometric flux of luminous infrared galaxies ($L_{8-1000\mu{\rm m}}>10^{11}~L_\odot$); their relation is based on all four \IRAS\ fluxes.

\begin{figure}[!ht]
\plotone{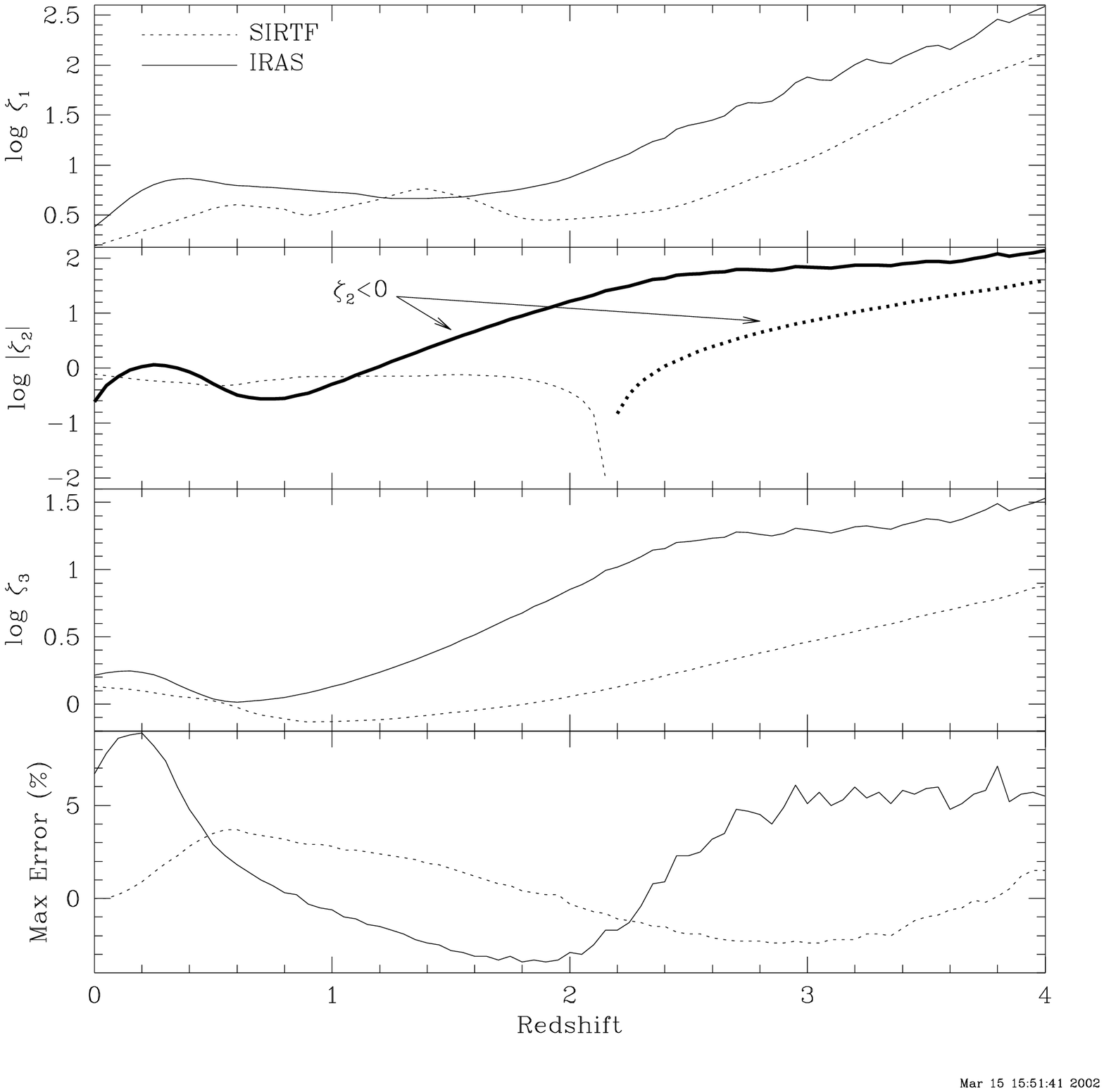}
\caption[] {\ A plot of how the coefficients of Equations~\ref{eq:LTIR_sirtf} and \ref{eq:LTIR_iras} vary with redshift.  The thick lines in the $\zeta_2$ panel indicate where the coefficients are negative.  The bottom panel displays the maximum percentage error encountered for the full suite of normal galaxy SED shapes; applying Equation~\ref{eq:LTIR_sirtf} or \ref{eq:LTIR_iras} to a particular normal galaxy SED model results in a discrepancy no greater than that displayed in the bottom panel.}  
\label{fig:L_TIR}
\end{figure}

It would be useful to derive a relation that is bolometric and applicable to a wider range of galaxy infrared luminosities.  Moreover, it should be based on \IRAS, \ISO, or \SIRTF\ fluxes.  A simple combination of \SIRTF\ Multiband Imaging Photometer (MIPS) fluxes recovers the total $3-1100$\m\ flux (TIR) for the full range of normal galaxy infrared SED shapes:
\begin{equation}
L_{\rm TIR}=\zeta_1\nu L_\nu(24\mu{\rm m}) + \zeta_2\nu L_\nu(70\mu{\rm m}) + \zeta_3\nu L_\nu(160\mu{\rm m})
\label{eq:LTIR_sirtf}
\end{equation}
where $[\zeta_1,\zeta_2,\zeta_3]=[1.559,0.7686,1.347]$ for $z=0$.  The coefficients are derived from a singular value decomposition solution to an overdetermined set of linear equations (Press et al. 1996).   Figure~\ref{fig:L_TIR} gives the coefficients for all $z$ out to four.  Equation~\ref{eq:LTIR_sirtf} matches the model bolometric infrared luminosity, for all model SED shapes, to better than 1\% at $z=0$, and to within 4\% for all redshifts smaller than four (see the bottom panel of Figure~\ref{fig:L_TIR}).  For archival purposes, the analagous \IRAS-based relation for normal galaxies is:
\begin{equation}
L_{\rm TIR}=\zeta_1\nu L_\nu(25\mu{\rm m}) + \zeta_2\nu L_\nu(60\mu{\rm m}) + \zeta_3\nu L_\nu(100\mu{\rm m})
\label{eq:LTIR_iras}
\end{equation}
where $[\zeta_1,\zeta_2,\zeta_3]=[2.403,-0.2454,1.6381]$ for $z=0$.  At $z=0$, the latter relation for \IRAS\ fluxes 
reproduces the model bolometric infrared luminosities to better than 1\% for most galaxies, though it does become inaccurate by about 7\% for the colder galaxies; the \IRAS\ 100\m\ bandpass does not extend to long enough wavelengths to effectively capture the coldest dust emission in nearby galaxies.

\subsection{The Submillimeter-to-Radio Ratio}

Efforts have been made to probe the utility of the submillimeter-to-radio flux ratio as a distance indicator (Lilly et al. 1999; Carilli \& Yun 1999; Blain 1999; Dunne, Clements \& Eales 2000; Barger, Cowie \& Richards 2000).  The technique is influenced by the effect that star-formation has on the emission in the far-infrared/submillimeter and in the radio: stars heat up dust grains which reprocess that energy as thermal emission; and stars that go supernova accelerate electrons throughout the interstellar medium, where they produce synchrotron emission (e.g. Helou \& Bicay 1993).  Since 850\m\ and 1.4~GHz lie on either side of the millimeter trough in normal galaxy (rest-frame) SEDs, and because the synchrotron emission increases with wavelength whereas the submillimeter dust emission falls with wavelength, the ratio \colord\ will increase with redshift.  As one can infer from the spectra in Figure~\ref{fig:seds}, the ratio is somewhat constrained for the suite of normal galaxies, and should systematically grow larger for increasing redshifts (at least up to $z\sim5$, before the far-infrared peak of the colder galaxies begins to enter the SCUBA 850\m\ bandpass).

\begin{figure}[!ht]
\plotone{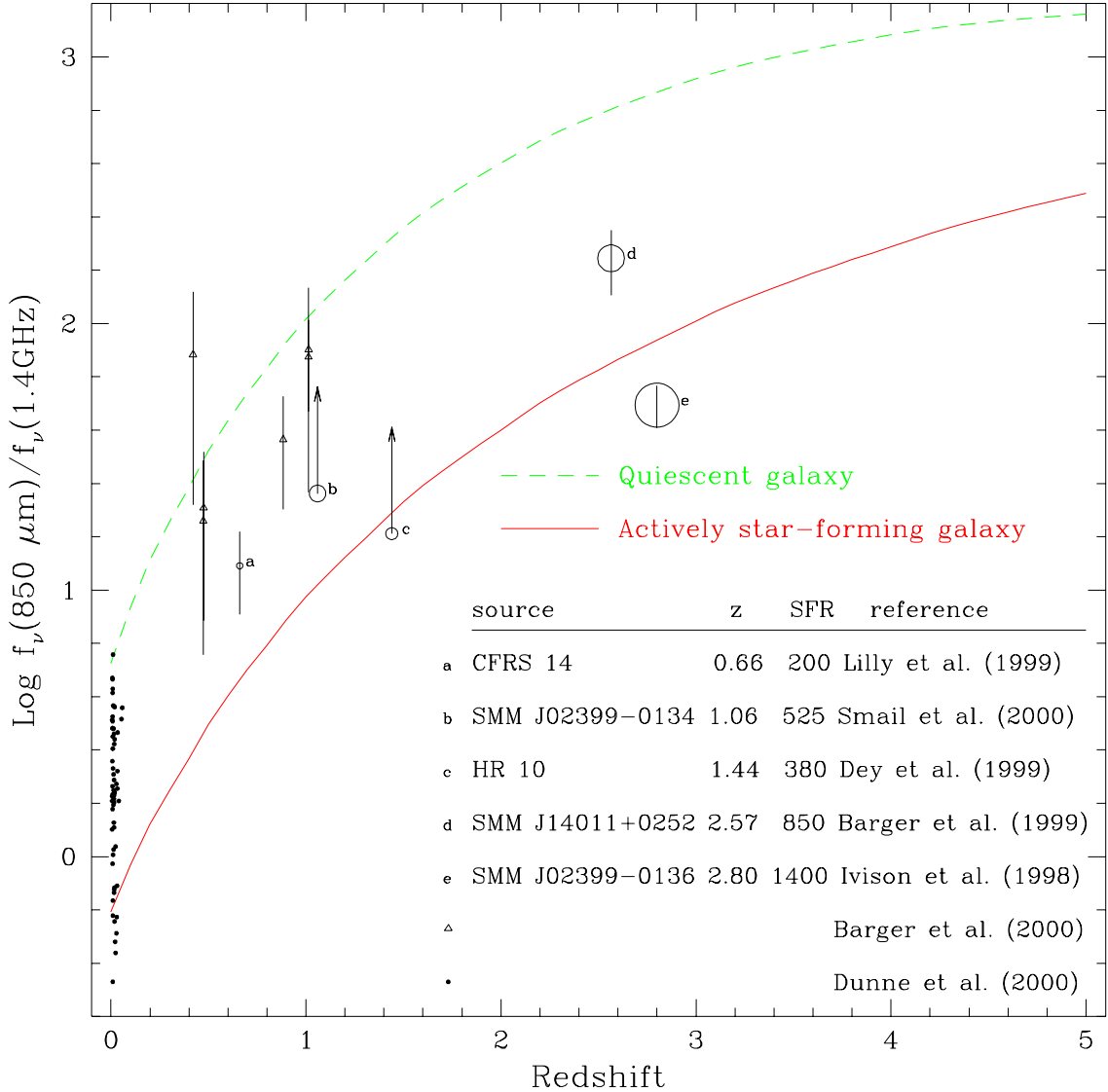}
\caption[] {\ The potential utility of the submillimeter-to-radio ratio as a redshift indicator.   The normal galaxy envelope is delineated by the predictions for a typical cold galaxy ($\alpha=1.06$) and a typical actively star-forming galaxy ($\alpha=2.31$).  The data included from Barger et al. (2000) are only those with signal-to-noise greater than 1.  For clarity, the error bars for the low redshift objects are not indicated.  For a few sources only the 4.85~GHz radio flux is available.  For these sources the 1.4~GHz flux has been estimated assuming $S_\nu \propto \nu^{-0.8}$.  Star formation rates ($M_\odot~{\rm yr}^{-1}$) are estimated by Yun \& Carilli (2001) for five higher redshift sources; the star formation rates for these galaxies are proportional to the size of the open circles.}
\label{fig:submm_radio}
\end{figure}

The envelope displayed in Figure~\ref{fig:submm_radio} gives the predicted normal galaxy range of the submillimeter-to-radio flux density ratio as a function of redshift.  Unless additional data on a given source can provide clues to its intrinsic infrared colors, the diversity of normal galaxy SEDs results in a rather broad envelope to the relation, and thus limits its ultimate utility.  For example, if log~\colord$\approx1$, then $0.2\lesssim z \lesssim 1$, and if log~\colord$\approx2$ then $1\lesssim z \lesssim 2.8$.  Note that the envelope is defined by models based on nearby normal galaxies, whereas the submillimeter sources are likely different in at least some important aspects.  For example, SMM~J02399-0136 appears to be an extremely active hyperluminous infrared galaxy ($L_{\rm FIR}\sim10^{13}~L_\odot$) with up to half of its luminosity deriving from an active galactic nucleus.  Its SED may not be well represented in our models since the most infrared luminous galaxies in the sample used for the models have $L_{\rm FIR} < 10^{12}~L_\odot$ and no obvious AGN activity (Dale et al. 2000).  This additional uncertainty as to the intrinsic SED of the high redshift sources can only add to the difficulty of using this technique to constrain redshifts.

\subsection{The Mid-Infrared-to-Submillimeter Ratio}
\label{sec:mir_submm}

An interesting and perhaps remarkable aspect of the figures displayed in Figure~\ref{fig:seds} is the approximately constant ratio between the mid-infrared and submillimeter fluxes; if far-infrared and submillimeter fluxes are normalized by the flux at 7\m, the ratios exhibit a narrow dispersion among normal galaxy model SEDs, generally in the 10--20\% range.  Table~\ref{tab:ratios} provides the ratios and their dispersions.  The dispersion is more like 30--50\% for actual data.  Combining the mid-infrared data from Dale et al. (2000) and the submillimeter from Dunne et al. (2000) for 15 normal galaxies, $\left<{f_\nu (6.75 \mu {\rm m})} \over {f_\nu (850 \mu {\rm m})}\right> \approx 1.5$ (with a dispersion of $\sigma \approx 0.8$).  For comparison, Haas et al. (2001), who use a similar ratio\footnote{Haas et al. used the continuum-subtracted 7.7\m\ peak flux.  See also Klaas et al. (2001) and Haas et al. (2002).} as a probe of mid-infrared extinction, find $\left<{f_\nu (7.7 \mu {\rm m})} \over {f_\nu (850 \mu {\rm m})}\right> \approx 4~(\sigma \approx 2)$  for a sample of 20 normal galaxies.  In a new study of five spatially-resolved galaxies, Haas et al. (2002) find a 30\% dispersion within each galaxy: $\left<{f_\nu (6.75 \mu {\rm m})} \over {f_\nu (850 \mu {\rm m})}\right> \approx 1.0~(\sigma \approx 0.3)$.

An obvious application of this mid-infrared-to-submillimeter constancy would be the ability to estimate submillimeter fluxes using only mid-infrared data.  Thus, if far-infrared data are also available, approximate dust masses are possible without even a measurement in the submillimeter.  Such a tool would be particularly useful in quickly constraining the properties of galaxies in future large mid-infrared extragalactic surveys like the {\it SIRTF Wide-area Infrared Extragalactic Survey} (Lonsdale 2001), the {\it Great Observatories Origins Deep Survey} (Dickinson 2001), and the {\it SIRTF First Look Survey} (Shupe et al. 1999).

\begin{figure}[!ht]
\plotone{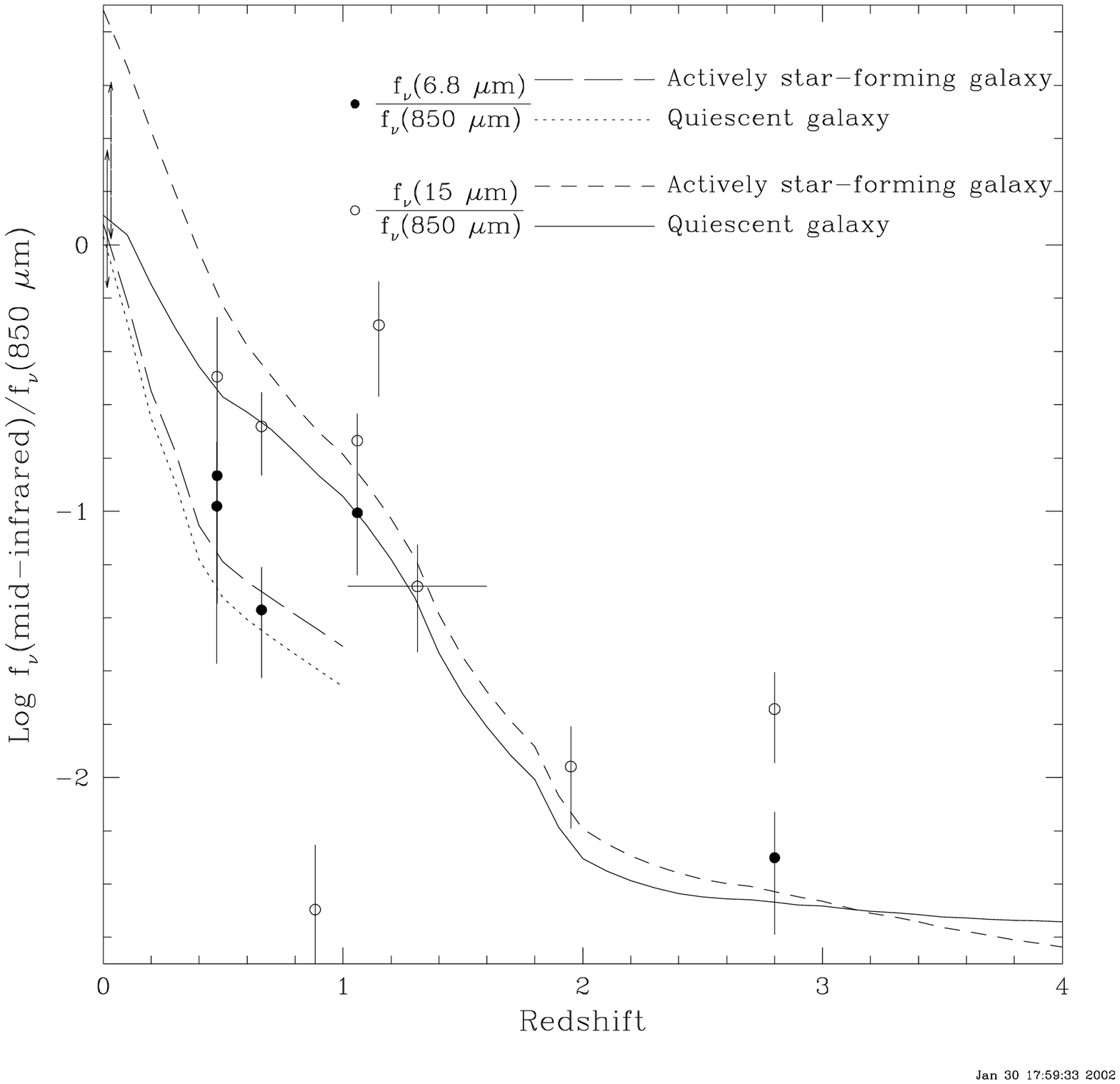}
\caption[] {\ Modeled mid-infrared-to-submillimeter ratios as a function of redshift.  Normal galaxy envelopes are given as in Figure~\ref{fig:submm_radio}, and the available literature data are also plotted as open (\colorc) and filled (\colorb) circles.  The observational data come from: HDF850.3~J123644.8+621304 @ $z_{\rm phot}\sim1.95$ (Hughes et al. 1998; Aussel et al. 1999); CFRS~14 @ $z=0.66$ (Lilly et al. 1999); SMM~J02399-0136 @ $z=2.80$ (Ivison et al. 1998; Smail et al. 2002); SMM~J02399-0134 @ $z=1.06$ (Smail et al. 2000; Smail et al. 2002); SMM~J21536+1742 @ $z\sim1.31$ (Smail et al. 2002); CUDSS~14.13 @ $z=1.15$ (Eales et al. 2000); and three sources from Barger et al. (2000).  The \colorb\ curves stop at $z=1$ since the SED model is only calibrated with \ISO\ down to $\lambda=2.5$\m.  The redshift zero ranges ($\pm1\sigma$) are indicated by arrows and stem from Dale et al. (2000) and Dunne et al. (2000).}  
\label{fig:submm_mir}
\end{figure}

Another application involves photometric redshifts.  Beyond the local Universe, the mid-infrared-to-submillimeter ratio should drop as the redshifted far-infrared peak approaches submillimeter wavelengths (see Figure~\ref{fig:seds}).  Figure~\ref{fig:submm_mir} displays two mid-infrared-to-submillimeter ratios as a function of redshift: \colorb\ and \colorc.  In contrast to the submillimeter-to-radio trends portrayed in Figure~\ref{fig:submm_radio}, the models predict the mid-infrared-to-submillimeter ratio to be fairly tight locally, and to remain so at higher redshifts.  Moreover, the rather steep profiles of the curves out to $z\sim1-2$ lend these ratios to be better suited as distance indicators at intermediate redshifts.  However, we caution that the suite of SED models are constructed to represent the average shapes of normal galaxies---the actual data can show large variations (e.g. see Figure~10 of Dale et al. 2001).  As noted above, the observed local dispersion is two to three times larger than what the models predict, so future high redshift may show the actual normal galaxy envelopes to be larger by a similar factor than as indicated in Figure~\ref{fig:submm_mir} (the data points in the figure may stem from extreme sources).

\section{Summary}

The phenomenological model for the infrared spectral energy distributions for normal star-forming galaxies presented by Dale et al. (2001) does not adequately predict galaxy submillimeter fluxes and far-infrared fluxes beyond 100\m.  In this work the empirical calibration is extended to submillimeter wavelengths; to better reproduce the far-infrared \ISO\ and submillimeter SCUBA data, the model's dust emissivity index is slightly changed as a function of the interstellar radiation field intensity.  The modification is such that cold dust is removed from the previous modeling of the most quiescent regions and added to the most actively star-forming region models.  Clearly, a power law distribution of dust mass over different heating environments, coupled with a variable dust emissivity, is not the only possible approach to modeling galaxy infrared fluxes.  Others have found that simple models involving only two thermal dust components and $\beta\sim2$ can adequately represent galaxy far-infrared/submillimeter SEDs (e.g. Dunne et al. 2001; Popescu et al. 2000).  Or perhaps a broken power law more accurately reflects the dust mass distribution.   Nevertheless, several recent findings point to a dust emissivity that depends on environment, and have led us to this approach.  Additional data from \SIRTF, spanning a wide variety of galaxy types and infrared wavelengths, should help to improve upon these various approaches and point to the most promising version.

The well-known far-infrared-to-radio correlation allows us to extend the spectral energy distributions to even longer wavelengths, and thus to explore the feasibility of the submillimeter-to-radio ratio as a redshift indicator.  Though the range of spectral shapes of normal galaxies unfortunately results in a rather broad submillimeter-to-radio envelope, the relatively tight mid-infrared-to-submillimeter ratio shows more promise as a redshift tool.  More data are needed, however, before any real confidence can be placed in the latter's redshift predictability.  Finally, we reach the important conclusion that the standard method for estimating dust mass needs substantial revision, especially for the coldest galaxies where the underestimate is a factor of ten.

\acknowledgements 
We would like to thank J. Brauher for sharing with us the data from his archival ISOLWS template study, S. Bianchi for providing his SCUBA map of NGC~6946, and W. Reach for helpful discussions.  \ISO\ is an ESA project with instruments funded by ESA member states (especially the PI countries: France, Germany, The Netherlands and the United Kingdom) and with the participation of ISAS and NASA.  This research has made use of the NASA/IPAC Extragalactic Database which is operated by JPL/Caltech, under contract with NASA.

\scriptsize
\begin{deluxetable}{rrc}
\tablenum{1}
\label{tab:fits}
\def\a{\tablenotemark{a}}
\def\p{$\pm$}
\tablewidth{240pt}
\tablecaption{Fitted Slopes of Data/Model Trends\a}
\tablehead{
\colhead{Wavelength} & \colhead{Before} & \colhead{After}
\\
\colhead{($\mu$m)}   & \colhead{}       & \colhead{} 
}
\startdata
SCUBA  850 & 1.23\p0.07 & 0.39\p0.07\\
ISOPHT 170 & 0.40\p0.09 & 0.20\p0.09\\
ISOLWS 170 & 0.51\p0.10 & 0.27\p0.11\\
ISOLWS 158 & 0.62\p0.08 & 0.43\p0.09\\
ISOLWS 145 & 0.59\p0.08 & 0.47\p0.08\\
ISOLWS 122 & 0.39\p0.06 & 0.37\p0.06\\
ISOLWS ~88 & 0.38\p0.11 & 0.39\p0.11\\
ISOLWS ~63 & 0.02\p0.11 & 0.02\p0.11\\
ISOLWS ~57 & 0.09\p0.09 & 0.11\p0.09\\
ISOLWS ~52 & $-$0.02\p0.12 & 0.03\p0.13\\
\enddata
\tablenotetext{a}{\footnotesize Derived from a least squares fit to log (flux observed/flux predicted by model) vs. log \IRAScolor, both ``before'' and ``after'' invoking the long-wavelength variable dust emissivity described by Equation~\ref{eq:beta_tweak}.}
\end{deluxetable}
\normalsize

\scriptsize
\begin{deluxetable}{ccccccccccc}
\tablenum{2}
\label{tab:budget}
\def\a{\tablenotemark{a}}
\tablecaption{Infrared Energy Budget}
\tablehead{
\colhead{log \IRAScolor} & \colhead{$\alpha$\a}        & 
\colhead{3-5 $\mu$m}     & \colhead{5-13 $\mu$m}     & \colhead{13-20 $\mu$m}        & 
\colhead{20-42 $\mu$m}   & \colhead{42-122 $\mu$m}   & \colhead{122-1100 $\mu$m}
\\
\colhead{}   & \colhead{}& \colhead{\%} & \colhead{\%} & \colhead{\%} & 
\colhead{\%} & \colhead{\%} & \colhead{\%}
}
\startdata
$+$0.10 & 1.09 & 0.5 & ~3.4 & 4.8 & 32.8 & 52.4 & ~6.1 \\
$+$0.00 & 1.41 & 0.8 & ~4.8 & 4.4 & 28.0 & 53.3 & ~8.7 \\
$-$0.10 & 1.64 & 1.3 & ~6.7 & 4.0 & 22.7 & 52.7 & 12.5 \\
$-$0.20 & 1.82 & 1.7 & ~8.9 & 3.8 & 17.8 & 50.8 & 16.9 \\
$-$0.30 & 2.00 & 2.2 & 11.2 & 3.7 & 13.6 & 47.9 & 21.5 \\
$-$0.40 & 2.22 & 2.7 & 13.4 & 3.7 & ~9.9 & 43.6 & 26.7 \\
$-$0.50 & 2.56 & 3.0 & 15.3 & 3.8 & ~7.4 & 38.3 & 32.2 \\
\enddata
\tablenotetext{a}{\footnotesize $\alpha$ is the index of the power law distribution described by Equation~\ref{eq:dMdU}.}
\end{deluxetable}
\normalsize

\scriptsize
\begin{deluxetable}{rccc}
\tablenum{3}
\label{tab:ratios}
\def\a{\tablenotemark{a}}
\def\b{\tablenotemark{b}}
\def\r{$\left< \log {f_\nu ({\rm MIR}) \over f_\nu ({\rm FIR})} \right>$}

\tablewidth{320pt}
\tablecaption{Model Mid-Infrared-to-Far-Infrared/Submillimeter Flux Density Ratios}
\tablehead{
\colhead{Wavelength} & \colhead{\r} & \colhead{Model Dispersion} & \colhead{${\rm Dispersion \over Mean}$}
\\
\colhead{($\mu$m)} & \colhead{} & \colhead{(1$\sigma$ in dex)\b} & \colhead{(\%; linear)}
}
\startdata
~148           & $-1.38\pm$0.01 & 0.053 & 11 \\
\SIRTF\ ~160\a & $-1.88\pm$0.01 & 0.053 & 12 \\
~191           & $-1.22\pm$0.01 & 0.029 & ~6 \\
~245           & $-0.99\pm$0.01 & 0.056 & 12 \\
~314           & $-0.72\pm$0.02 & 0.074 & 16 \\
~403           & $-0.39\pm$0.02 & 0.079 & 17 \\
~518           & $-0.04\pm$0.02 & 0.076 & 16 \\
~665           & $+0.34\pm$0.01 & 0.067 & 15 \\
SCUBA 850\a    & $+0.09\pm$0.01 & 0.054 & 12 \\
~854           & $+0.70\pm$0.01 & 0.049 & 11 \\
1100           & $+1.07\pm$0.01 & 0.055 & 12 \\
\enddata
\tablenotetext{a}{\footnotesize These data are calculated by convolving the model SEDs with the \ISO\ 6.75\m, the \SIRTF\ 160\m, and the SCUBA 850\m\ filter profiles.  The remaining rows of data are from model monochromatic flux density ratios, including the peak model flux density at 7.7\m\ (no continuum subtraction has been performed for the 7.7\m\ flux density).}
\tablenotetext{b}{\footnotesize The model ``population'' dispersion---the $1\sigma$ value from 25 different SED shapes that span the normal galaxy range.}
\end{deluxetable}
\normalsize

\end{document}